# Dynamic control of active droplets using light-responsive chiral liquid crystal environment

Vanessa Jirón[a,b], Mojtaba Rajabi[a,c], Hao Wang[a,b], Oleg D. Lavrentovich[a,b,c,*]

[a]Advanced Materials and Liquid Crystal Institute, Kent State University, Kent, Ohio 44242, USA
[b]Materials Science Graduate Program, Kent State University, Kent, Ohio 44242, USA
[c]Department of Physics, Kent State University, Kent, Ohio 44242, USA

*Author to whom correspondence should be addressed: olavrent@kent.edu

ORCID: Vanessa Jirón (https://orcid.org/0000-0003-2145-9921), Mojtaba Rajabi (https://orcid.org/0000-0002-4738-449X), Hao Wang (https://orcid.org/0000-0001-9109-6017), Oleg D. Lavrentovich (https://orcid.org/0000-0002-0128-0708)

## ABSTRACT

Microscopic active droplets are of interest since they can be used to transport matter from one point to another. In this work, we demonstrate an approach to control the direction of active droplet propulsion by a photoresponsive cholesteric liquid crystal environment. The active droplet represents a water dispersion of bacterial *Bacillus subtilis* microswimmers. When placed in a cholesteric, a surfactant-stabilized active droplet distorts the local director field, producing a point defect-hedgehog, with fore-aft asymmetry, and allows for the chaotic motion of the bacteria inside the droplet to be rectified into directional motion. When the pitch of the cholesteric confined in a sandwich-like cell is altered by light irradiation, the droplet trajectory realigns along a new direction. The strategy allows for a non-contact dynamic control of active droplets trajectories and demonstrates the advantage of orientationally ordered media in control of active matter over their isotropic counterparts.

**Introduction**

Active matter is comprised of living or artificial out-of-equilibrium units capable of transforming the stored energy or the energy harnessed from the environment into directional motion.[1–3] One example is an emulsified spherical droplet that becomes motile due to interactions in the emulsion. Most self-propelled droplets in isotropic environments are driven into motion by the Marangoni effect. The mechanism of their propulsion is either by chemical reactions or solubilization at the surface of the droplet, both of which cause gradients in the surface tension of the dispersing medium and break the symmetry of the droplet, causing it to self-propel.[2,4,5] Complex motion is generally observed in these systems, but ballistic motion is also achievable.[4] However, due to the inherent lack of ordering in isotropic media, the control of the direction of the flows of the droplets is a challenge,[6–10] which is why liquid crystals (LCs) are a convenient alternative.[11]

An active droplets emulsion in a LC could be, for example, a spherical water droplet containing a dispersion of swimming bacteria that is placed in a thermotropic nematic LC confined between two plates with planar molecular orientation. When the surface alignment of LC molecules is perpendicular to the droplet's surface, the mismatch between the local radial molecular orientation around the droplet and planar alignment at the plates results in a topological point defect, called a hyperbolic hedgehog (HH), Fig.1a,b.[12–14] The director field of the droplet-HH pair is fore-aft asymmetric. We describe this asymmetry by a vector $\mathbf{P}$ directed from the droplet center towards the HH core. The flows created by the randomly swimming bacteria inside the droplet transfer through the interface into the nematic surrounding. The polar asymmetry of molecular orientations around the sphere-HH pair rectifies these chaotic flows into a directional flow of the nematic outside the droplet. As a result, the droplet propels unidirectionally along $\mathbf{P}$.[15] In addition to enabling and directing the locomotion of active droplet, the nematic environment adds a useful effect of levitation: the droplet is repelled from the top and bottom plates of a sandwich-like cell by the elastic response of the nematic to different anchoring conditions at the droplet's surface and at the plates.

As demonstrated previously, the trajectory of the active droplet could be controlled by surface patterning of the director,[15] by applying an electric field to realign the director, or by using a laser beam to locally melt the nematic into an isotropic phase.[16] These methods of control are either static (as in the case of photopatterning[15]) or require a complicated design (such as patterned electrodes or precise focusing of the laser beam[16]). There is thus a need to develop a method to dynamically control the trajectories of active droplets with a simpler design that can be applied from a distance. We propose to address this challenge by using a light-responsive cholesteric (Ch) LC as the medium for the dispersed active droplets.

In a Ch, the director is arranged in a helical supramolecular structure, which is characterized by the pitch $p$, defined as the distance over which the director twists by $360°$. The pitch is sensitive to chemical composition. In particular, the pitch of a Ch mixture containing chiral molecules with photosensitive groups, such as azobenzene moieties, could be tuned by an exposure to light which causes *trans*-to-*cis* isomerization and thus changes molecular interactions responsible for the equilibrium $p$.[17–20] Imagine that an active droplet is placed in a Ch confined between two glass plates, one of which is rubbed to produce a unidirectional alignment along an axis $x$. The second plate allows the director to glide on it. The thickness $d$ of the cell is close to $p$ or less than $p$. The Ch helicoidal axis is perpendicular to the bounding plates. The equilibrium coordinate $z_0$ of the droplet's center along the helicoidal axis is defined by the balance of gravity and elastic repulsion from the bounding plates.[21] The dipole **P** is along the direction that makes an angle

$$\varphi = 2\pi z_0 \, /p \qquad (1)$$

with the $x$-axis. If the pitch changes under light irradiation, $p \to p_{\mathrm{irr}}$, while $z_0$ remains constant, the dipole **P** must change its orientation to $\varphi_{\mathrm{irr}} = 2\pi z_0/p_{\mathrm{irr}}$ which is generally different from $\varphi$. Since the droplet moves along **P**, one expects the change of the pitch to change the direction of propulsion in the $xy$-plane of the cell.

Figures 1b,c illustrate the concept of realignment of **P** for a droplet of a diameter $2R$ that is much smaller than the Ch pitch $p$. The director field in the equatorial plane in this case

is close to the one in the nematic, Fig.1b. If the pitch is changed, but the equilibrium position $z_0$ remains constant, this quasinematic director field and the vector **P** realign along a new direction. For example, if the new pitch is $p + p/4$, **P** realigns by 90°, Fig.1c. Figures 1d and 1e illustrate the idea further by showing how **P** realigns when the Ch handedness is reversed. Therefore, a tunable Ch pitch can be used as a steering parameter to control the propulsive direction of an active droplet.

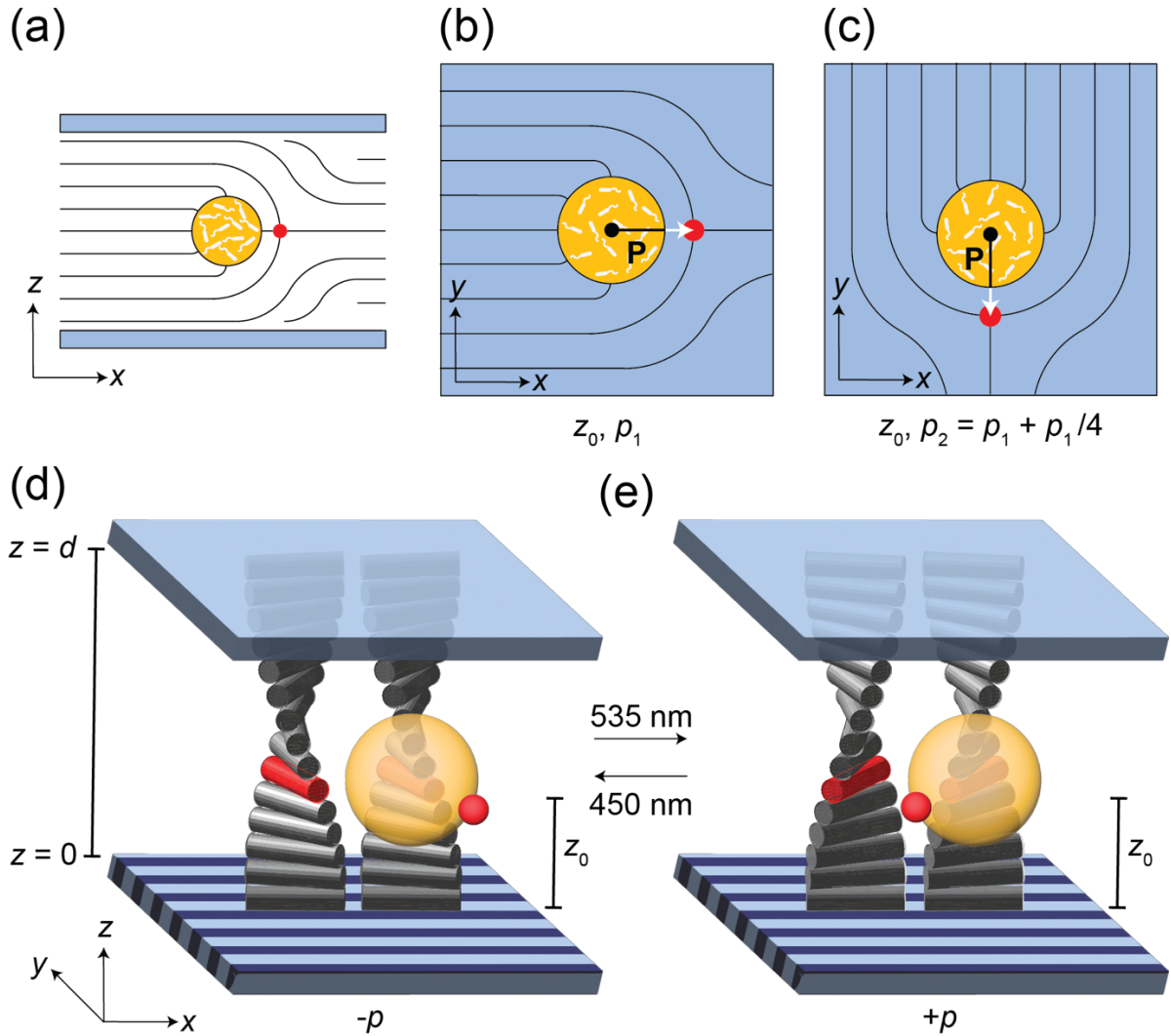

FIG. 1. Director configuration of a liquid crystal around an active droplet. (a) Director configuration of a nematic around an active droplet of a diameter $2R$ with homeotropic surface anchoring inside a cell with unidirectional planar anchoring along $x$ on both plates.

The red circle represents the hyperbolic hedgehog core. (b) View of the cell from the top plate. The vector $\mathbf{P}$ characterizing the fore-aft asymmetry of the director field around the droplet is directed from the center of the droplet towards the hyperbolic hedgehog core (red circle). The angle $\varphi$ is defined between the projection of $\mathbf{P}$ onto the cell's $xy$-plane and the rubbing direction $x$. The equatorial director pattern in a Ch with $p \gg 2R$ is similar to the nematic one. (c) Realignment of the director pattern and $\mathbf{P}$ caused by the change of the Ch pitch. (d,e) Realignment of $\mathbf{P}$ caused by the reversal of the Ch handedness in a cell with planar alignment at the bottom plate and tangentially degenerate anchoring at the top plate.

To implement the above idea, we use a Ch in which the pitch could be controlled by light within the visible part of the spectrum, to avoid potentially harmful effects of irradiation on the bacteria. Namely, the azobenzene-based chiral dopant exhibits a reversible isomerization from *trans* to *cis* on exposure to light of the wavelength 530 nm, and from *cis* to *trans* when exposed to light of the wavelength 440 nm. The photoisomerization causes a change in the pitch and handedness of the Ch material.[17] In general, winding and unwinding of the Ch helix realign the polar axis $\mathbf{P}$ of the droplets along a new azimuthal direction, which means that the visible light provides the dynamic and non-contact means to control the trajectories of active droplets.

**Results**

Below we first describe how the light irradiation changes the Ch pitch $p$ and then how the change of $p$ controls the trajectories of active droplets. Since the orientation of the dipole depends on $p$, Fig.1b-e, we use Ch mixtures with different concentrations of chiral dopant to determine how the overall trajectories of the droplet could be controlled and characterize the dynamics of reorientation after subsequent changes in the wavelength of light irradiation.

**A. Light-induced changes of the Ch pitch**

The equilibrium pitch and the helicity of the Ch are determined by the $\theta$-cell method.[22,23] The $\theta$-cell is used only for the pitch measurements and is not used for the experiments on controlled propulsion of active droplets. The $\theta$-cell is formed by two parallel plates, the bottom one with planar anchoring and the top one with circular anchoring.[24,25] The conflicting anchoring conditions lead to the formation of a disclination line. For a nematic, the disclination appears at $0°$ with respect to the rubbing direction. For a left-handed Ch, the disclination realigns clockwise (when viewed from the top plate side) towards the II and IV quadrants at an angle $\delta < 0$ measured with respect to the rubbing direction. For a right-handed Ch, the disclination realigns counterclockwise (when viewed from the top plate side) towards the I and III quadrants at $\delta > 0$.[22] The pitch $p$ is determined as $p = 2\pi d/\delta$, where $d < p$ is the cell thickness.[22] The measured pitch values of two Ch mixtures of BrDAB (Fig.2b) in 5CB (Fig.2a) are displayed in Table I. To ensure that the value of $p$ corresponds to the equilibrium state, each measurement was performed after 10 min stabilization.

(a)                    (b)

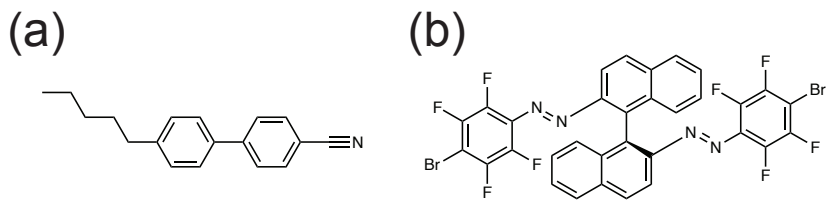

FIG. 2. Chemical structures of the components of the Ch. (a) 4-$n$-pentyl-4'-cyanobiphenyl (5CB) liquid crystal host and (b) light-sensitive chiral dopant ($S$)-2,2'-bis(($E$)-(4-bromo-2,3,5,6-tetrafluorophenyl)diazenyl)-1,1'-binaphthalene (BrDAB).

Table I. Cholesteric pitch of 5CB + BrDAB mixtures at different wavelengths of steady irradiation.

| Molar fraction of BrDAB | Pitch at 535 nm / μm | Pitch at 450 nm / μm |
|:---:|:---:|:---:|
| $4.9 \times 10^{-4}$ | $273 \pm 1$ | $-262 \pm 1$ |
| $8.8 \times 10^{-4}$ | $201 \pm 1$ | $-201 \pm 1$ |
| $12.1 \times 10^{-4}$ | $140 \pm 1$ | $-156 \pm 1$ |

The chiral dopant undergoes *trans*-to-*cis* isomerization on exposure to green 535 nm light, and the reversed process on exposure to blue 450 nm light.[17] The *trans* configuration of BrDAB produces a left-handed Ch, while the *cis* configuration of BrDAB produces a right-handed Ch. The transition from one isomer to another requires an unwinding of the helix, passing through the nematic state, and twisting the helix in an opposite direction.[17] Note that our experiments use two extreme guiding conditions, when only one wavelength is used at a time, to illustrate that the trajectories depend on the *trans/cis* content. One can also control the dynamics by two beams of different wavelength and variable intensities, to achieve smooth variations of the pitch.

## B. Reorientation dynamics of active droplets by light irradiation in a Ch

The Ch cell induces planar anchoring of the director on the bottom plate $z = 0$ and azimuthally degenerate tangential anchoring on the top plate, $z = d$, Figs.1d and 1e. Degenerate tangential anchoring allows the material to adjust the pitch after every irradiation step. The anchoring at the bottom plate is set planar by unidirectional rubbing, so that when the pitch changes, the polar vector **P** of asymmetric director field around the droplet rotates. In the absence of planar anchoring, the director can glide at the bounding plate, which could prevent **P** from realignment. When a colloidal particle with homeotropic anchoring of the director is placed in a nematic cell, it causes the formation of topological defects in the surrounding LC:[12] a surface disclination loop, called a Saturn ring (SR),[26] or a point defect, called a hyperbolic hedgehog (HH).[12] The SR configuration is of a quadrupolar symmetry and does not cause the propulsion of an active droplet.[15] In a nematic, the relative stability of the equatorial SR and HH depends on the cell thickness-to-droplet diameter $d/2R$ ratio: the SR is stable in shallow cells, $d/2R \rightarrow 1$,[27,28] while the HH is stable when $d/2R > 1$.[29] In a Ch, the director field is more complex than in nematics and depends also on the pitch-to-droplet diameter $p/2R$ ratio.[30–32] High values of $\frac{p}{2R} \gg 1$ produce textures similar to that one of a nematic, Fig.1a,b,c. For $\frac{p}{2R} \ll 1$, the appearance of HH is highly unlikely, since the strongly twisted planar Ch structure "flattens" the director lines to be parallel to the bounding plates. Instead, the condition $\frac{p}{2R} \ll 1$ favors a disclination loop that encircles the sphere approximately $2R/p$ times, being a twisted analog of the SR.[32,33] As demonstrated experimentally and by numerical simulations[30] for $p/2R \approx 1-2$, the HH still forms in the equatorial plane of the spherical colloid, but in addition to it, there are also nonsingular disclination loops enclosing the top and bottom parts of the sphere. Since the director is continuous in these loops, they are hardly visible under the microscope.[30]

It is expected that since the ratio $p/2R$ controls the stability of the hedgehog defect, it should be important also in the efficiency of flow rectification around the active droplet. When $\frac{p}{2R} \ll 1$, the dipolar asymmetry of the director field around a droplet diminishes and the rectification of the chaotic flows produced by the bacteria inside the droplet is no longer efficient in the surrounding Ch. When $\frac{p}{2R} \gg 1$, the dipolar structure with the HH should rectify the flows well, but the realignment angle might be diminished. To avoid these limiting cases, we explored different $\frac{p}{2R}$, by adjusting the pitch and the droplet size. In one case, a strongly twisted Ch, with $p_{535} = -p_{450} = 201$ μm, contains a droplet of a diameter $2R = (134 \pm 2)$ μm, which yields $\frac{|p|}{2R} \approx 1.5$, Fig.3. A case of an even smaller $\frac{|p|}{2R} \approx 1.0$ is presented in Supplementary Fig.1. Figure 4 illustrates the behavior with a large $\frac{|p|}{2R} \approx 3.7 - 3.8$, in which case $p_{535} = 273$ μm, $p_{450} = -262$ μm, and $2R = (72 \pm 4)$ μm. An intermediate case of $\frac{|p|}{2R} \approx 2.1$ is presented in Supplementary Fig.2. In agreement with the prior results by Trivedi *et al.* [30] for a solid sphere of diameter $\sim 10$ μm, the Ch environment in all cases creates a dipolar structure with a hedgehog defect, suitable for the rectification of active flows and propulsion of the droplet, see Figs.3,4 and Supplementary Fig.1. Furthermore, fluorescence confocal microscopy of the vertical cross-section of the cells[34] shows that the active droplets levitate in the Ch bulk, Supplementary Fig.3, as discussed for inanimate colloids by Pishnyak *et al.*[21,35] The equatorial plane of the large droplet, Supplementary Fig.3a is close to the midplane of the cell, $z = d/2$. The details of the propulsion and trajectory control, however, depend on the ratio $\frac{|p|}{2R}$, which can be used as an optimization parameter depending on the tasks in the controlled steering of active droplets. We first describe the strongly twisted Ch with $\frac{|p|}{2R} \approx$ 1.5, Fig.3.

### 1. *Strongly twisted Ch, $\frac{|p|}{2R} \approx 1.5$*

The dipolar nature of the director field around the large active droplet, $2R = (134 \pm 2)$ μm, is enhanced by a small satellite droplet trapped at the core of the HH, Fig.3(a). The hedgehog point defect cores carry a large elastic energy of director distortions. In the presence of foreign particles, this energy can be reduced by attracting the particle to the core, thus removing the strongly distorted liquid crystal.[36] In the present case, the hedgehog core attracts a small droplet which facilitates the observations and makes the polar axis **P** and the angle $\varphi$ it makes with the $x$-axis clearly defined, Fig.3a.

It is important to stress that the presence of the satellite droplet at the HH core is not the reason for the propulsion of the principal droplet. As explained in Discussion, the satellite droplet merely reduces the high core energy of the hedgehog. The principal active droplets show the same dynamic behavior even if there are no satellite droplets, as shown in Supplementary Figures 1 and 2 and Supplementary Movies 2 and 3. The propulsion is enabled by the fore-aft asymmetry of the director field around the principal droplet and this asymmetry persists regardless of the presence or absence of the satellite droplet, Fig.1. The mechanism of propulsion is in the rectification of the chaotic bacteria-triggered flows by this fore-aft asymmetric director field. This mechanism is different from the nonreciprocal "predator-prey" interactions between two oil-in-water droplets which engage in oil exchange and Marangoni flows, as described by Meredith *et al.*[10] In addition, the polar axis **P** of the fore-aft asymmetry of the director field around the active droplet is controlled by the far-field director, which enables the control of the propulsion direction by this far field.

The situation is also different from the other LC-involving propulsion phenomenon described by Krüger *et al.*,[5] which involves 5CB droplets dispersed in water with an ionic surfactant. The droplets dissolve with time since the concentration of the ionic surfactant is above the critical micelle concentration. In this dynamic regime, the 5CB droplets propel. Propulsion is observed when 5CB is in the isotropic phase and also in the nematic phase. The isotropic 5CB droplets move along rectilinear trajectories while the nematic 5CB droplets show a curling motion. In both cases, propulsion is caused by self-sustained gradients of the surfactant concentration at the droplets' surface, which produces Marangoni flows and drives the droplets; the coupling of the director inside the nematic droplets brings an additional symmetry-breaking mechanism that converts ballistic motion into a curled one. The phenomenon is different from the present case of bacterial active droplets. First, although our system does contain a surfactant (lecithin), its presence is not the reason for the propulsion. If the bacteria are deprived of oxygen for a long time and stop swimming, there is no propulsion of the droplets. Second, the environment in which the droplets propel in our case is a Ch liquid crystal, rather than isotropic water. This orientationally ordered environment produces director distortions around a surfactant-covered droplet which are of a polar symmetry, as specified by the vector **P**. If the liquid crystal environment is heated into the isotropic phase, the active droplets do not propel;[15] they experience a random Brownian motion.

Whenever the wavelength of light irradiation is changed, the Ch adjusts its pitch, while the droplet adjusts the angle $\varphi$ of the polar axis **P**, Fig.3a. The director field around the droplet experiences a strong reorganization, which involves shifts of the hedgehog core away from the droplet and transient appearance of nonsingular disclinations, Supplementary Movie 1. The readjustments are relatively slow, taking about 6-7 min, Fig.3d,e, which is much longer than the characteristic time 75 s for the light-induced isomerization of BrDAB.[37]

Once the reorganization is complete, the direction of propulsion changes, as expected, Fig.3b,c. **P** rotates clockwise going from frames 1 to 2 and frames 3 to 4 when irradiation with 535 nm is switched to the 450 nm irradiation, Figure 3a and Supplementary Movie 1. The result is consistent with the clockwise twist described for the *cis*-to-*trans* transition caused by this irradiation change. The droplet rotates counterclockwise when the switch from 450 nm to 535 nm causes a *trans*-to-*cis* transition of the light-sensitive chiral BrDAB.

The propulsion of the active droplet is ballistic and uniform once the Ch structure readjusts to the new irradiation regime. In this steady regime, the velocity direction and magnitude $ds/dt$ are constant, Fig.3(b,c,d,e). Here, $s$ is the distance by which the droplet progresses every 60 s.

Comparison of Fig.3a to Fig.3b also shows that the droplet propulsion direction is noticeably different from the direction of $\mathbf{P}$, in contrast to the case of a nematic environment, in which these two directions coincide.[15] For example, the droplet moves along the azimuthal direction $\varphi_{d,535} \approx 77°$ under 535 nm irradiation and along $\varphi_{d,450} \approx -67°$ under 450 nm irradiation, Figs.3b,e. These values are different from the average orientation angles of $\mathbf{P}$ in the steady state, $\langle \varphi_{535} \rangle = 44°$ and $\langle \varphi_{450} \rangle = -46°$, Fig.3e. Another important difference is that the propulsion speed of droplets with a diameter $2R \approx 130\,\mu m$ in a nematic environment is about $1\,\mu m\,s^{-1}$, while the strongly twisted Ch with $\frac{|p|}{2R} \approx 1.5$, Fig.3 and $\frac{|p|}{2R} \approx 1.0$, Supplementary Fig.1, reduces the speed to about $0.2\,\mu m\,s^{-1}$. The apparent reason for these differences between the Ch with $\frac{|p|}{2R} \approx 1$ and a nematic is a more complicated structure of the defects around the droplet in a Ch environment, which contains additional non-singular disclinations revealed by numerical simulations[30] and observed as weakly light-scattering strings in Supplementary Movies 1 and 2. The relatively strong structural twist and the presence of additional disclinations reduce the fore-aft asymmetry of the overall director configuration around the active droplet, which diminishes the efficiency of flow rectification and sets the direction of propulsion different from that of $\mathbf{P}$. The twist and disclinations also produce an additional viscous drag on the moving active droplet caused by the reconstruction of the director field. Hence, when $\frac{|p|}{2R} \lesssim 1$, the effectiveness of the rectification of chaotic bacterial flows is diminished.

As demonstrated in the next section, increasing the parameter $\frac{|p|}{2R}$ allows one to significantly increase the speed of self-propulsion, Fig.4.

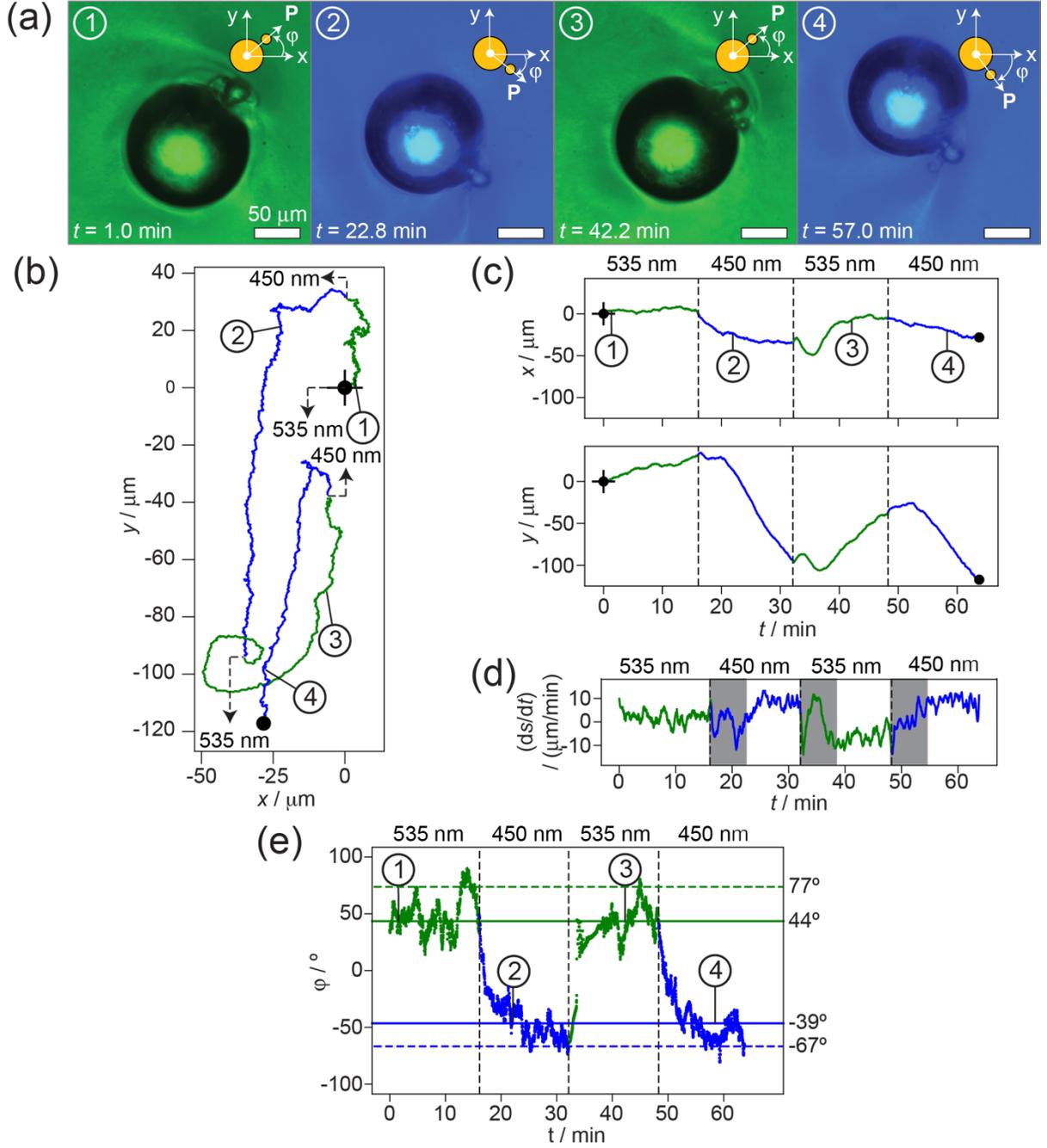

FIG. 3. Active droplets trajectories by light irradiation in a strongly twisted Ch. (a) Optical microscope view of an active droplet with a diameter $2R = (134 \pm 2)\,\mu m$ suspended in a Ch

mixture of 5CB with BrDAB at molar fraction at $8.8 \times 10^{-4}$ upon irradiation with green 535 nm light and blue 450 nm light in four different time frames: 1) $t = 1.0$ min, 2) $t = 22.8$ min, 3) $t = 42.2$ min, 4) $t = 57.0$ min. The inset is a schematic representation of how $\varphi$ is measured. (b) Trajectory of the droplet. (c) Time dependence of the displacements of the droplet in the $x$- and $y$-axes. (d) Time derivatives of the displacements ($s = \sqrt{x^2 + y^2}$) of the droplet. (e) Time dependence of the angle $\varphi$ between **P** and the $x$-axis. The solid lines are the median $\varphi$ for each wavelength calculated from the observed $\varphi$ in the optical microscope; the dashed lines are the median $\varphi$ calculated from the droplet trajectories.

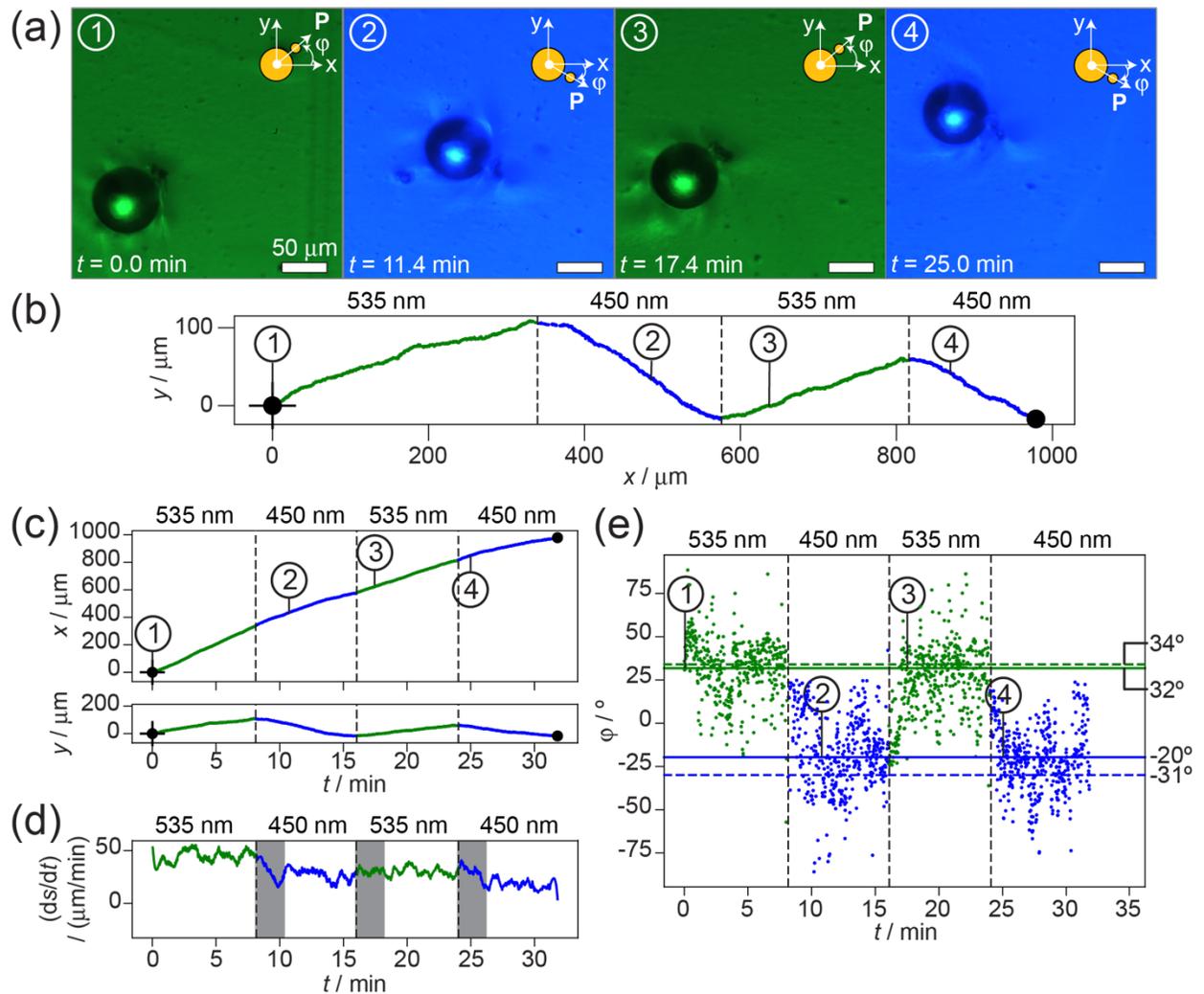

FIG. 4. Active droplets trajectories by light irradiation in a weakly twisted Ch. (a) Optical microscope view of an active droplet with a diameter $2R = (72 \pm 4)$ $\mu$m suspended in a Ch mixture of 5CB doped with BrDAB in molar fraction $4.9 \times 10^{-4}$ upon irradiation with green 535 nm light

and blue 450 nm light in four different time frames: 1) $t = 0.0$ min, 2) $t = 11.4$ min, 3) $t = 17.4$ min, 4) $t = 25.0$ min. The inset is a schematic representation of how $\varphi$ is measured. (b) Trajectory of the droplet. (c) Time dependence of the displacements of the droplet in the $x$- and $y$-axes. (d) Time derivatives of the displacements ($s = \sqrt{x^2 + y^2}$) of the droplet. (e) Time dependence of the angle $\varphi$ between **P** and the $x$-axis. The solid lines are the median $\varphi$ for each wavelength calculated from the observed $\varphi$ in the optical microscope; the dashed lines are the median $\varphi$ calculated from the droplet trajectories.

2. *Weakly twisted Ch,* $\frac{|p|}{2R} \approx 3.7 - 3.8$

In a weakly twisted Ch with $\frac{|p|}{2R} \approx 3.7 - 3.8$, the clockwise and counterclockwise reorientations of the droplet upon changing the wavelength of light irradiation, Fig.4a, are faster than the reorientations in the strongly twisted Ch, Fig.4b,c,d,e. The uniform motion and constant speed of propulsion are recovered after 2 min of irradiation change; the change in the wavelength of irradiation does not cause the appearance of additional disclinations, Supplementary Movie 4, in contrast to the case with a smaller $\frac{|p|}{2R} \approx 1.5$ described above and $\frac{|p|}{2R} \approx 1.0$ in the Supplementary Movie 2. The droplet trajectory is along $\varphi_{d,535} \approx 30°$ under 535 nm irradiation and along $\varphi_{d,450} \approx -30°$ under 450 nm irradiation, Fig.3b,e. These directions coincide with the direction of **P**, Supplementary Movie 2. This behavior is the same as in the nematic environment.[15]

The higher ratio $\frac{|p|}{2R}$ yields a higher speed of active droplets. In Figure 4d, the speed of a droplet of the diameter $2R = (72 \pm 4)$ μm fluctuates in the range $(0.3 - 0.9)$ μm s$^{-1}$, which is comparable to the speed $0.4$ μm s$^{-1}$ of similarly sized droplets in the nematic environment[15] and noticeably faster than the speed $0.2$ μm s$^{-1}$ of the larger droplet $2R = (134 \pm 2)$ μm in the case $\frac{|p|}{2R} \approx 1.5$ in Fig.3. The difference between the weakly and strongly twisted Ch is even more striking if one accounts for the fact that the large droplets move faster than the small ones in a nematic.[15]

The equilibrium value of the angle $\varphi$ at each wavelength of steady irradiation depends on both the droplet diameter and the pitch. The angular range of reorientations $\Delta\varphi_d = \varphi_{d,535} - \varphi_{d,450}$, could be tuned by changing the concentration of the chiral dopant BrDAB. At high concentrations, the director twists by a higher $\varphi$ for a given cell thickness, Eq. (1), which allows for larger degree of angular steering of an active droplet. In this respect, a strongly twisted Ch allows for a broader range of trajectory realignments, as clear from Fig.3(b), in which $\Delta\varphi_d \approx 180°$. For a weakly twisted Ch, the trajectory realignment range is smaller, $\Delta\varphi_d \approx 55°$, Figure 4b.

If one uses Equation (1) for the droplet in the weakly twisted Ch, the expected values of $\varphi$ are $\varphi_{535} = 48°$ and $\varphi_{450} = -50°$, so that $\Delta\varphi_d \approx 100°$, which is larger than the experimental $\Delta\varphi_d \approx 55°$. Equation (1) assumes that the ideal Ch helix is not distorted by the droplet and that the hedgehog geometry is the same as in nematics. However, previous studies[30–32] demonstrate that the director fields around a sphere with homeotropic anchoring in a Ch and a nematic are very different. Our data for the strongly twisted Ch also demonstrate that the direction of propulsion is not parallel to **P**, Fig.3b,e, which would make $\Delta\varphi_d$ different from the one expected on a simple argument of Equation (1). Furthermore, the value $\Delta\varphi_d$ calculated from Equation (1) assumes that the droplet is located at the exact same $z_0$-level when irradiated with two different light beams. However, since the cholesteric pitch is somewhat different in the two cases, $p_{535} = 273\ \mu m$ and $p_{450} = -262\ \mu m$, the elastic interaction[21] of the droplet with the bounding plates and thus the $z_0$-locations might be different in the two cases (also in the intermediate unwound nematic state), which would result in a different $\Delta\varphi_d$. Other potential reasons such as distortions by viscous friction during self-propulsion, should not affect the spherical shape of the droplets much since the capillary number $Ca = \frac{\eta v}{\sigma}$ is small. Here $\eta$ and $v$ are the viscosity and velocity of the moving droplet, and $\sigma$ is the interfacial tension between the two fluids. For the viscosity of DSCG 13% $\eta \approx 7\ kg\ m^{-1}\ s^{-1}$.[38] and $v = 0.6\ \mu m\ s^{-1}$, assuming that $\sigma$ is in a very broad range from 1 to 100 mN m$^{-1}$, the capillary number is in the range $Ca = 10^{-6} - 10^{-8}$, much smaller than 1.

The reorientation of the droplets in the weakly twisted Ch after subsequent changes in the wavelength of light irradiation is reproducible, as the equilibrium $\varphi_{535}$ is the same in frames 1 and 3, and the equilibrium $\varphi_{450}$ is the same in frames 2 and 4 in Fig.4a. These values remain the same after two cycles of irradiation change.

**Discussion**

Propulsion and trajectory of active droplets in the Ch environment are controlled by a number of factors, such as the size of the droplet, the number of microswimmers in them, the cholesteric pitch, and the $z$-coordinate of the droplet center. We discuss these parameters and associated limits below.

The droplet size is limited from below by three factors. The obvious requirement is that the droplet diameter is larger than the length of the bacterium, which in the case of *B. subtilis* places the lower limit at about 10 μm. Second, the droplets should be sufficiently large to maintain a perpendicular surface anchoring at the LC-Terrific Broth interface and strong fore-aft asymmetry of the director field around them, which defines the efficiency of flow rectification. The surface anchoring energy of a droplet scales approximately as $WR^2$, where $W$ is the anchoring coefficient, while the elastic energy of distortions around the droplet scales as $KR$.[39] As a result, droplets smaller than $R < K/W$ do not perturb much the uniform surrounding director field. For a typical $W = 10^{-6}$J/m$^2$ and $K = 10$ pN, the critical droplet radius is $R_c \sim \frac{K}{W} \sim 10$ μm. Droplets of a radius smaller than $R_c$ do not show a strong fore-aft asymmetry, do not rectify the chaotic flows induced by bacteria, and thus would not self-propel efficiently. Increasing $W$ and decreasing $K$ would make the limit $R_c$ smaller.

It is important to stress that the scaling $WR^2$ of the surface anchoring energy and $KR$ of the elastic energy controls also the properties of the small satellite droplet that sometimes, Fig.3, but not always, Fig.4 and Supplementary Figs.1 and 2, is found at the core of the hyperbolic hedgehog. This droplet is not needed to induce the self-propulsion of the prime larger drop. It is thus very different from the small droplets in the predator-prey scenario described by Meredith *et al.*[10] The satellite droplet in Fig.3 is of a small radius $R_s \approx 12$ μm, close to the critical estimate $R_c \sim \frac{K}{W} \sim 10$ μm above. These small droplets form during stirring of the liquid crystal-bacterial water dispersion mixture in the process of emulsification. The small droplets tend to find the locations with the strongest director distortions. In Fig.3, the droplet replaces the strong director distortions of an elastic energy $KR_s$ at the hedgehog core, at the expense of the anchoring energy $WR_s^2$. If $R_s < K/W$, such a replacement is energetically preferred, as explained by Voloschenko *et al.*[36] Such a satellite does not change the surrounding LC much and does not satisfy the strict homeotropic anchoring conditions at its surface, precisely because $WR_s^2 < KR_s$. In particular, it hardly influences the fore-aft asymmetry and topological transformations around the prime droplet, such as the hyperbolic hedgehog-to-Saturn ring, as illustrated by the comparison of Fig.3 case with the satellite and Fig.4, Supplementary Figs. 1, 2 cases without the satellite. The topological charge of the hyperbolic hedgehog in the presence of the satellite droplet is still -1, as in the case when the satellite droplet is not present at the core.

The upper limit on the droplet's size is apparently set by the thickness of the chamber and by the requirement that the local director environment around the droplet is not strongly twisted and preserves a sufficient fore-aft asymmetry of the director field. This condition yields the upper limit $2R \approx |p|$, which is corroborated by the data in Fig.3 and Supplementary Fig.1 that show a slow self-propulsion when $2R \approx |p|$.

Number of bacteria is the third factor limiting the droplets size from below. As established experimentally by Rajabi *et al.*,[15] droplets smaller than about 30 μm do not show self-locomotion, apparently because the small number of bacteria in them cannot produce strong internal flows. Additionally, the propulsion speed of the droplets also increases with increasing concentration of bacteria, with the optimum being in the range $(1.6 - 2.4) \times 10^{25}$ cells m$^{-3}$.[15]

The pitch of the cholesteric matrix could be independently adjusted to the requirements such as $2R < |p|$ by changing the composition, e.g., the molar fraction $x_{\mathrm{BrDAB}}$ of a chiral dopant, since $|p| \propto x_{\mathrm{BrDAB}}^{-1}$. For example, the pitch changes from $\approx 270$ μm when $x_{\mathrm{BrDAB}} = 4.9 \times 10^{-4}$, Fig.4, to $\approx 140$ μm when $x_{\mathrm{BrDAB}} = 12.1 \times 10^{-4}$, Supplementary Fig.S1. Since the bacterial active droplets capable of self-propulsion should be larger than about $30$ μm, we conclude that the shortest pitch of interest should be about $30$ μm. The upper limit on pitch is set by the desired steering angle $\Delta\varphi_{\mathrm{d}}$. If the steering angle for a droplet of a diameter $\approx 70$ μm is desired to be about $55°$, as in Fig.4(b), then the pitch should be $\approx 270$ μm or less. A smaller pitch would increase $\Delta\varphi_{\mathrm{d}}$ but simultaneously decrease the speed of propulsion, compare Fig.4 to Fig.3.

Levitation of active droplets is an added benefit of using an LC environment instead of an isotropic fluid. The droplet levitates since gravity is counterbalanced by the elastic repulsion of the director distortions around the droplet from the bounding plates. [21,35] In a nematic, the elastic repulsion force scales as $R^4$. Since gravity force scales as $R^3$, large droplets levitate better, which means that their equatorial plane is closer to the middle plane of the confinement cell. This effect is also relevant for the Ch environment, as confirmed by the fluorescence confocal microscopy textures[34] of the vertical cross-sections of Ch in a $d = 150$ μm cell containing a large, $2R = 90$ μm, and a small, $2R = 40$ μm, active droplets, Supplementary Fig.3. The large droplet levitates at a higher $z$, close to the midplane of the cell, $z = d/2$.

An enormous advantage of an LC environment is that its director controls the locomotion direction of active units in it. As already stated, the fore-aft director asymmetry enabling the active droplet's propulsion is caused by the surrounding far-field director of the liquid crystal medium. This far-field director can be designed as rectilinear or spatially and temporarily varied by several approaches, such as photopatterning of surface interactions,[40] application of the external electric field,[16] or by light irradiation, which allows one to control the propulsion direction. In the presented example, the photocontrol of the cholesteric pitch changes the in-plane direction of active droplet propulsion, so that the control can be called two-dimensional. However, the liquid crystal environment allows one to add dynamic control along the third dimension, perpendicular to the cell's plane. For this purpose, one can use a liquid crystal medium with a negative dielectric anisotropy and apply a direct current (DC) field across the cell (by using two transparent electrode coatings at the bounding glass plates). As demonstrated by Lazo *et al*,[41] a colloidal particle in such an environment can be moved along the normal to the cell by linear electrophoresis, thus adding the third dimension to the control of propulsion. Such a control is hard to achieve with isotropic environment.

**Conclusion**

This work demonstrates that blanket light irradiation can be used to control the trajectories of active droplets. When suspended in a light-sensitive cholesteric, active droplets rotate counterclockwise or clockwise, depending on the wavelength of irradiation, which causes either *trans*-to-*cis* or *cis*-to-*trans* isomerization of the photosensitive chiral azobenzene additive. The isomerization causes reversal of the cholesteric handedness and changes the cholesteric pitch. As a result, the active droplet with dipolar asymmetry of the Ch director around it, realigns its polar axis **P** and moves in a new direction specified by the wavelength of irradiation. An important factor controlling the propulsion is the pitch-to-diameter ratio $\frac{|p|}{2R}$. Relatively small values $\frac{|p|}{2R} \approx 1.5$, corresponding to strongly twisted Ch, yield a higher range of trajectory realignment, which reaches about $180°$. However, the propulsion speed and the transient regimes of structural reorganization are slow. In contrast, weakly twisted Ch with a higher $\frac{|p|}{2R} \approx 3.7 - 3.8$ offers a propulsion speed that is similar to the one in a nematic environment and reduces the reconstruction time of the director during irradiation to about 2 min. These features are related to a higher degree of polar asymmetry in the case of a higher $\frac{|p|}{2R}$.

The presented approach allows one to dynamically control the trajectories of active droplets using light as a non-contact stimulus, with a simple experimental design. In this work, the driving force causing propulsion of active droplets is the activity of bacterial microswimmers. However, the approach to control the propulsion of the droplets by a cholesteric environment is not limited by bacteria: any interior or surface-active flows of droplets would produce similar propulsion and re-direction by the cholesteric environment if the director field outside the droplet is of a dipolar asymmetry.

**Methods**

**A. Materials**

The thermotropic nematic LC 4-*n*-pentyl-4'-cyanobiphenyl (5CB, Fig.1(a)) (Synthon) serves as the host. The chiral dopant (*S*)-2,2'-bis((*E*)-(4-bromo-2,3,5,6-tetrafluorophenyl)diazenyl)-1,1'-binaphthalene (BrDAB, Fig.1(b)) is synthesized as described previously.[17] Hexane (Sigma-Aldrich, >99%) is used as a solvent for BrDAB. Deionized water (Thermo Scientific Barnstead RO purifier, 18.2 MΩ-cm) is used for all aqueous solutions. The medium for bacteria growth is composed of terrific broth (TB) powder (Sigma-Aldrich) and glycerol (Sigma Life Science, >99%) in water. The surfactant *L-α*-phosphatidylcholine (lecithin) from egg yolk (Sigma Life Science, >99%) added to the LC imposes homeotropic anchoring of the director at the surface of the active droplets. Disodium cromoglycate (DSCG) (Alfa Aesar, 98%) is added to the aqueous dispersion of bacteria to increase the viscosity of the medium for better momentum transfer through the droplet-LC interface.[15] The solutions of dye Brilliant Yellow (BY) (Aldrich Chemistry, 70%) in dimethylformamide (DMF) (Sigma-Aldrich, 99.8%) are used for patterned photoalignment of the LC. The nematic MAT-03-382 (Merck), chiral dopant *S*-(+)-2-octyl 4-(4-hexyloxybenzoyloxy)benzoate (S811) (TCI, >90%) and fluorescent dye *N,N'-bis*(2,5-di-tert-butylphenyl)-3,4,9,10-perylenedicarboximide (BTBP) (Aldrich Chemistry, >85%) are used for fluorescent confocal polarized microscopy.

**B. Preparation of light-sensitive cholesteric liquid crystal**

BrDAB is dissolved in hexane (0.13% by weight). This solution is mixed with 5CB to obtain a homogeneous mixture. The BrDAB +hexane+5CB solutions are heated in a reduced-pressure oven (0.1 bar) at 50 ℃ for 24 h to evaporate the hexane completely. The molar fractions of BrDAB in the resulting Ch mixtures are $x_{\text{BrDAB}} = 4.9 \times 10^{-4}$, $x_{\text{BrDAB}} = 8.8 \times 10^{-4}$ and $x_{\text{BrDAB}} = 12.1 \times 10^{-4}$.

## C. Preparation of active droplets

*Bacillus subtilis* (strain 3280) is grown in lysogeny broth (LB) agar plates (Teknova) at 35 ℃ for 24 h. One colony of *B. subtilis* is subsequently grown in a TB medium (47.6 g/L TB powder, 8 mL L⁻¹ glycerine, deionized water) at 35 ℃ for 9 h in a shaking incubator (Boekel Scientific). The bacteria are separated from their growth medium by centrifugation and suspended in the DSCG solution to achieve a final concentration of $2.4 \times 10^{16}$ cells m⁻³. Bacteria with the DSCG solution (2 $\mu$L of this solution) are suspended in the light-sensitive Ch (50 $\mu$L) and stirred in a vortex mixer to obtain a suspension of droplets of various sizes.

## D. Ch cells assembly with active droplets

The active droplets emulsion is introduced into cells (of a thickness $h \approx 150\,\mu$m) with a rubbed polyimide (PI2555) coating on the bottom substrate and an unrubbed polystyrene coating on the top substrate, Fig.2(a). The cell thickness was chosen to allow a wide range of droplet sizes to enter the cell, as big droplets move faster than small droplets.[15] The unidirectionally rubbed polyimide PI2555 yields planar anchoring parallel to the $x$-axis.[42] Unrubbed polystyrene yields an azimuthally degenerate tangential alignment.[43] The polystyrene coating is produced by spin-coating (3000 rpm, 30 s) a solution of polystyrene beads in chloroform (Sigma-Aldrich, >99%) (5% by weight) over clean glass substrates and baking for 1 h at 90 ℃ to completely evaporate the solvent. The sides of the cells are sealed with epoxy glue after the filling. Experiments with active droplets are done within 2 h of sealing the cell.

## E. Preparation of $\theta$-cell

The $\theta$-cells are used to establish the pitch dependence on light irradiation.[22–25] BY is dissolved in DMF with constant stirring. The solution (0.5% by weight) is filtered, then spin-coated (3000 rpm, 30 s) over clean glass substrates and baked for 1 h at 90 ℃. The circular pattern of the dye in the coated substrate is achieved by a 10 min projection of a plasmonic metamask in a photopatterning setup.[44,45] These substrates are used for the top of the cell. The bottom substrate is coated with a rubbed polyimide PI2555 layer. Cells of a thicknesses $18\,\mu$m are filled with the light-sensitive Ch by capillary action. The pitch dependence on the wavelength of irradiation is measured in the steady regime allowing 10 min of the pitch relaxation.

## F. Optical microscopy

The dynamics of active droplets are studied under an inverted microscope (Nikon Eclipse TE2000-U) with a xenon light source (Asahi Spectra LAX-C100). Color filters are used to select the wavelength of light irradiation at wavelengths $\lambda = 535$ nm (bandwidth 50 nm, Chroma Technology) and 450 nm (bandwidth 10 nm, Asahi Spectra). The intensity of the light source is set at its maximum before inserting the filters and yields values of intensity of 9.7 mW cm$^{-2}$ and 1.4 mW cm$^{-2}$ after inserting the 535 nm and 450 nm filters, respectively. The room temperature is 22 ℃. A level is used to check there is no tilt of the microscope stage. Videos and pictures are recorded with a camera (Emergent Vision Technologies HS-20000C) connected to the microscope; video and image analysis is done with the softwares ImageJ and Maplesoft.

## G. Fluorescent confocal polarized microscopy (FCPM)

The $z$-coordinates of the droplets are characterized by fluorescent confocal polarized microscopy (Olympus Fluoview) with an argon laser (wavelength 488 nm) light source (NEC Corporation, model GLG30808).[46] The mixture for the FCPM observations is composed of DSCG droplets suspended in a low birefringence nematic MAT-03-382 doped with the fluorescent dye BTBP at 0.0025 % by weight and the chiral dopant S811 at 0.061% by weight (pitch $\sim\!150\,\mu$m). The cell (of thickness $h \approx 150\,\mu$m) has rubbed polyimide (PI2555) coating on the bottom substrate and an unrubbed polystyrene coating on the top substrate. Observations are made with the rubbing direction perpendicular to the polarizer axis of the FCPM.

**Data availability**

The data that supports the results herein presented are included within the text and the Supplementary Information. All other data that support the plots within this paper and other findings of this study are available from the corresponding author upon reasonable request.

**Acknowledgments**

This work is supported by NSF grant DMR-2215191.

**Author Contributions**

V.J. performed the experiments and analyzed the data. V.J. and M.R. designed the experimental setup. H.W. synthesized the BrDAB material. V.J. and O.D.L. wrote the manuscript with inputs from all coauthors. O.D.L. conceived and supervised the project.

**Competing interests**

All authors declare no competing interests.